\documentclass[aps,prl,twocolumn,showpacs,preprintnumbers,amsmath,amssymb,superscriptaddress]{revtex4}
\usepackage{graphicx}
\usepackage{dcolumn}
\usepackage{bm}

\begin{document}

\title{Mapping the Evolution of Optically-Generated Rotational Wavepackets\\in a Room Temperature Ensemble of D$_2$}
\author{W. A. Bryan}
\email{w.bryan@ucl.ac.uk}%
\affiliation{Department of Physics and Astronomy, University
College London, Gower Street, London WC1E
6BT, UK}%
\affiliation{Central Laser Facility, Rutherford Appleton
Laboratory, Chilton, Didcot, Oxon OX11 0QX, UK}
\author{E. M. L. English}
\affiliation{Department of Physics and Astronomy, University
College London, Gower Street, London WC1E 6BT, UK}
\author{J. McKenna}
\affiliation{Department of Pure and Applied Physics, Queen's
University Belfast, Belfast BT7 1NN, UK}
\author{J. Wood}
\affiliation{Department of Physics and Astronomy, University
College London, Gower Street, London WC1E 6BT, UK}
\author{C. R. Calvert}
\affiliation{Department of Pure and Applied Physics, Queen's
University Belfast, Belfast BT7 1NN, UK}
\author{R. Torres}
\affiliation{Blackett Laboratory, Imperial College London, Prince
Consort Road, London SW7 2BW, UK}
\author{I. C. E. Turcu}%
\affiliation{Central Laser Facility, Rutherford Appleton
Laboratory, Chilton, Didcot, Oxon OX11 0QX, UK}
\author{J. L. Collier}
\affiliation{Central Laser Facility, Rutherford Appleton
Laboratory, Chilton, Didcot, Oxon OX11 0QX, UK}
\author{I. D. Williams}%
\email{i.williams@qub.ac.uk}%
\affiliation{Department of Pure and Applied Physics, Queen's
University Belfast, Belfast BT7 1NN, UK}
\author{W. R. Newell}%
\email{w.r.newell@ucl.ac.uk}%
\affiliation{Department of Physics and
Astronomy, University College London, Gower Street, London WC1E
6BT, UK}
\date{\today}

\begin{abstract}
A coherent superposition of rotational states in D$_2$ has been
excited by nonresonant ultrafast (12 femtosecond) intense (2
$\times$ 10$^{14}$ Wcm$^{-2}$) 800 nm laser pulses leading to
impulsive dynamic alignment. Field-free evolution of this
rotational wavepacket has been mapped to high temporal resolution
by a time-delayed pulse, initiating rapid double ionization, which
is highly sensitive to the angle of orientation of the molecular
axis with respect to the polarization direction, $\theta$. The
detailed fractional revivals of the neutral D$_2$ wavepacket as a
function of $\theta$ and evolution time have been observed and
modelled theoretically.
\end{abstract}

\maketitle

The importance of being able to enforce spatial order on an
initially random ensemble of molecules has been recognized since
the discovery of steric effects in chemical reactions
\cite{Taft56}. Traditional `brute force' techniques employing
strong DC fields \cite{Loe90, Wu94} have recently yielded to new,
more subtle and yet more powerful and versatile techniques using
intense laser systems \cite{Seid99, Sta03}. In particular intense
femtosecond pulses have been successfully used to align molecules
along an axis for application in areas such as the study of
fragmentation dynamics \cite{Dool03, Pero03}, high harmonic
generation \cite{Ita04, Kan05} and the creation of single cycle
laser pulses \cite{Kal02}. In this case the alignment of a
molecule (or spatial ordering of an ensemble of molecules) evolves
temporally following the creation of a rotational wavepacket by a
linearly polarized laser pulse of far shorter duration than the
natural period of rotation.

In this Letter, we report on a detailed ultrafast study of the
temporal evolution of such a rotational wavepacket in the neutral
deuterium molecule D$_2$. The wavepacket is created impulsively by
a 12 femtosecond laser pulse, with temporal and angular ordering
probed at some later time with a similar pulse that initiates
sequential double ionization (SI) i.e. D$_2$ $\rightarrow$ D$_2^+$
$\rightarrow$ D$^+$ + D$^+$ \cite{Tong04}. Such high resolution
measurements represent the state-of-the-art in ultrafast molecular
physics in the high rotational frequency limit as dictated by this
most fundamental and theoretically tractable of molecules.

It is known that an intense linearly polarized laser pulse
interacting with an ensemble of molecules generates a degree of
spatial alignment from a random ensemble \cite{Seid99, Sta03}. The
only condition for this phenomenon is that the molecular
polarizability is anisotropic. Thus, the induced dipole moment
created in the interaction with the electric field of the laser
generates a torque that causes the axis of maximum polarizability
to librate around the polarization vector of the laser field.
Quantum mechanically, the process is described as a sequence of
Rabi-type cycles accompanied by the exchange of two quanta of
angular momentum between the molecule and the laser field. If the
aligning pulse duration is much greater than the rotational
period, the system evolves adiabatically to the the so-called
pendular states, which dissipate as soon as the laser pulse passes
\cite{Fried95}. However, when the pulse duration is much shorter
than the rotational period, a superposition of rotational states
is created that outlives the laser pulse, and the phases in the
rotational wavepacket continue to evolve \cite{Seid99}. As a
consequence of the quantization of the $\it{J}$ states, a periodic
dephasing and rephasing of the wavepacket occurs in the field-free
regime. At integer and half-integer multiples of
(2$\it{Bc}$)$^{-1}$ (where $\it{B}$ is the rotational constant and
$\it{c}$ is the speed of light in vacuum) a revival of the
rotational wavepacket is expected, at which point the ensemble
exhibits a significant degree of alignment, with the molecular
axes parallel to the polarization vector of the initial pulse. The
relative phases of the $\it{J}$ states also produces antialignment
in the ensemble, whereby the molecular axes are orientated to lie
on or near the plane normal to the initial pulse polarization.
Alignment and antialignment are well distinguished temporally. As
a consequence of the nuclear spin statistics, homonuclear systems
can also present partial revivals at 1/4 and 3/4 of the rotational
period.

Previous observations of impulsive alignment have been restricted
by the non-availability of laser pulses sufficiently shorter than
the rotational periods. However, rotational wavepackets have been
previously observed in heavy many-electron systems (N$_2$
\cite{Dool03}, O$_2$ \cite{Span04, Lee04}, CO$_2$ \cite{Ren04,
Xu06}, CS$_2$ \cite{Xu06}, I$_2$ \cite{Ros01} etc.), where
rotations occur with periods on the picosecond time scale. The
rotational revival in D$_2$ occurs at (2$\it{Bc}$)$^{-1}$ = 558
fs, requiring an aligning pulse of the order of tens of
femtoseconds in duration. Dynamic alignment of D$_2$ using 10 fs
pulses has recently been demonstrated by Lee $\it{et~al}$
\cite{Lee06}, however, these observations were of limited temporal
resolution and the angular evolution of the rotational wavepacket
was not measured.

In the present work, the spatio-temporal evolution of the D$_2$
rotational wavepacket has been simulated following the procedure
described in \cite{Torres05}. Briefly, a thermal ensemble of rigid
rotors is considered, each of which gives rise to a superposition
of rotational states, $|\Psi_{\it{J_i M_i}}\rangle$ = $\sum_{J
\geq \mid M_i \mid}$ $F_{J_i J}$(t)$|\it{J M_i}\rangle$ in which
$F_{J_i J}$($\it{t}$) are the time-dependent complex coefficients
(note that $\it{M_i}$ is conserved in a linearly polarized field).
These coefficients are calculated by numerically solving the
time-dependent Schr$\ddot{o}$dinger equation, and are used to
calculate the thermal average of the degree of alignment and the
angular evolution of the ensemble.

Due to the bosonic character of D$_2$, the total wavefunction of
the molecule must be symmetric under inversion. The electronic
ground state is symmetric, and the nuclei can form six symmetric
(ortho) and three antisymmetric (para) nuclear wavefunctions. This
restricts the $\it J$ states populated: para-D$_2$ occupies only
odd $\it{J}$-states, and ortho-D$_2$ occupies only even
$\it{J}$-states. This results in a 2 : 1 weighting (ortho-D$_2$ :
para-D$_2$) in a thermal ensemble. For a room temperature sample
we expect the following initial populations (in parenthesis) for
the different $\it{J}$ states: $\it{J}$ = 0 (0.185), 1 (0.208), 2
(0.386), 3 (0.112), 4 (0.0899), 5 (0.0128) and 6 (0.00522).

\begin{figure}
\includegraphics[width=180pt]{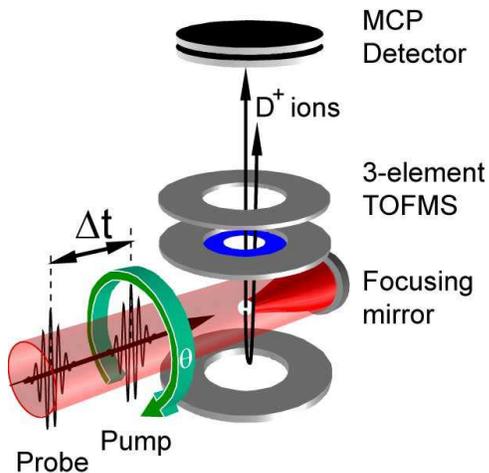}\\
\caption{The experimental configuration used to observe rotational
wavepackets in D$_2$ imprinted on the D$^+$ rapid sequential
ionization (SI) yield (D$_2$ $\rightarrow$ D$_2^+$ $\rightarrow$
D$^+$ + D$^+$) as measured with a time-of-flight mass spectrometer
(TOFMS). The time delay between the pump and probe pulses is
$\Delta$t, and the angle $\theta$ between pump and probe
polarization directions is varied by rotating the pump
polarization. As shown, $\theta$ = 0.}\label{fig1}
\end{figure}

The experiment was carried out using 800nm, 0.4 mJ, 10 fs pulses
at 1 kHz repetition rate. The pulses were split by a 4 $\mu$m
thick pellicle beamsplitter. A co-linear Mach-Zehnder
interferometer configuration was used to delay the resulting 70
$\mu$J, 12 fs probe pulse with respect to the 73 $\mu$J, 12 fs
pump pulse with 300 attosecond resolution \cite{McK06}. A
half-wave plate mounted in one arm of the interferometer allowed
the pump polarization to be rotated through an angle $\theta$ with
respect to the probe polarization, which was fixed parallel to the
spectrometer axis. In comparison to the core-annulus method
\cite{Lee06}, the present technique does suffer greater energy
loss in the split pulses. However, this is compensated by the
temporal consistency of the spatial overlap at focus as a function
of time delay over a far greater range. Furthermore the present
technique allows polarization control. These features are
essential to the present study.

\begin{figure}
\includegraphics[width=220pt]{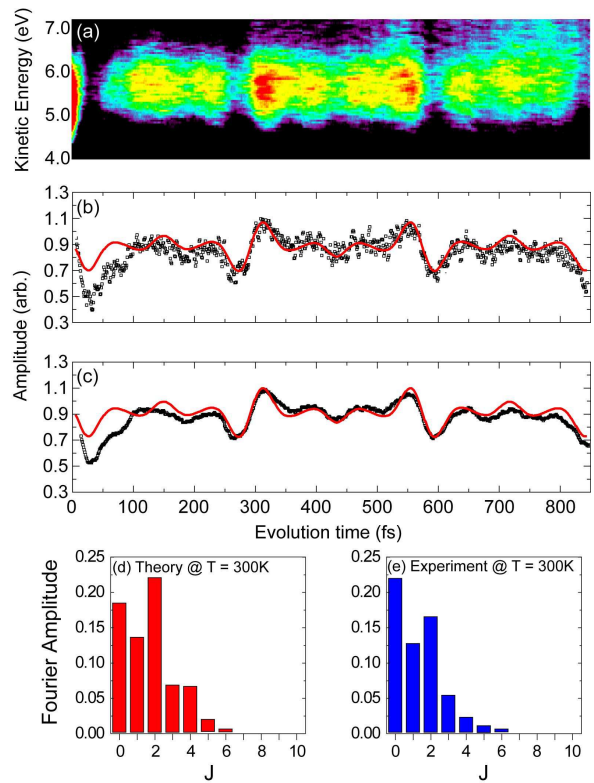}\\
\caption{(a) False colour representation of the D$^+$ rapid
sequential ionization (SI) yield (D$_2$ $\rightarrow$ D$_2^+$
$\rightarrow$ D$^+$ + D$^+$) as a function of pump-probe delay
(thus wavepacket evolution time) $\Delta$$\it{t}$ and ion kinetic
energy. (b) Integrated SI yield compared to theoretically
predicted influence of an impulsively excited rotational
wavepacket in D$_2$ (red). Throughout, $\theta$ = $\pi$/2, see
Fig. 1, and the step in $\Delta$$\it{t}$ = 1 fs. (c) Integrated SI
yield following an 11-point smoothing. The 1/4, 1/2 and 3/4
revivals are more clearly observed. Fourier Transform of the
theoretical simulation and experimental SI yield are shown in (d)
and (e) respectively. The frequency components present in the
rotational wavepacket are beats between rotational states
$\Delta\it{J}$ and $\Delta\it{J}$+2. The Fourier amplitudes thus
return a measure of rotational population.}\label{fig2}
\end{figure}

Following transmission through a fused-silica window into an
ultra-high vacuum chamber, the pump-probe pulses are reflection
focused with a silver-coated spherical mirror capable of
supporting the bandwidth of our laser pulses without introducing
group-velocity dispersion, as shown in Fig. 1. The laser focus is
situated in the source region of a tightly apertured (250 $\mu$m)
ion time-of-flight mass spectrometer (TOFMS) \cite{McK06}. The
aperture serves two purposes: to limit the angular acceptance
($\leq$ 2 degrees for a 5.5 eV D$^+$ ion), and to dramatically
limit the field of view of the spectrometer. By observing only the
central 250 $\mu$m `slice' of the $\it{f}$/5 focus, the instrument
is sensitive only to those molecules exposed to a narrow intensity
range. As a consequence, spatial integration over the focal volume
is obviated, and the behaviour of the rotational wavepacket is
clearly observed. This results in a remarkably high resolution
measurement of the rotational wavepacket as compared to earlier
observations.

Fig. 2(a) shows the measured rapid sequential ionization D$^+$
yield in false colour as a function of pump-probe delay
($\Delta$$\it{t}$, also referred to as evolution time) and D$^+$
ion kinetic energy; here, $\theta$ = $\pi$/2. We are interested in
the modulation in the SI yield 0 $\leq$ $\Delta$$\it{t}$ $\leq$
800 fs, the signature of the revival of a rotational wavepacket in
the D$_2$ molecule. Importantly, in Fig. 2, the orthogonality of
the aligning pump pulse to the TOFMS (and probe polarization) axis
results in the initial maximal alignment of the ensemble producing
a minimum in the SI signal as the majority of the molecules are
aligned perpendicular to the detector.

The solid circles in Fig. 2(b) represent the experimentally
obtained high kinetic energy D$^+$ yield from the rapid SI of
D$_2$ at small internuclear separation for $\theta$ = $\pi$/2. The
half- (280 fs) and full-revivals (560 fs) are clearly apparent in
the results of both the integrated SI yield and the theoretical
calculation, with excellent agreement observed between experiment
and theory. At small delays 0 $\leq$ $\Delta$$\it{t}$ $\leq$ 100
fs, the discrepancy between experiment and theory is the result of
the temporal wings of the pump and probe pulses overlapping to
produce elliptical/circular polarization depending on
$\Delta$$\it{t}$. When $\Delta$$\it{t}$ $\geq$ 200fs, the
time-dependent structure of the SI yield is purely the result of
the rotational wavepacket.

\begin{figure}
\includegraphics[width=240pt]{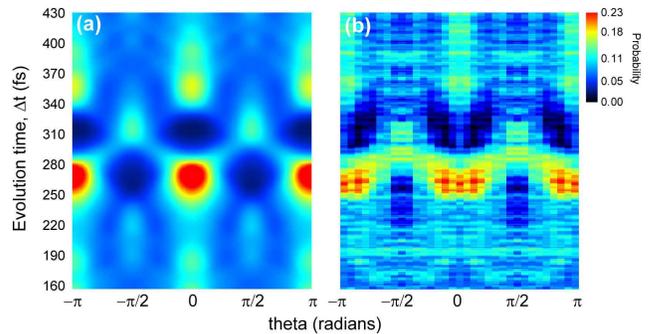}\\
\caption{(a) Simulated `quantum carpet' \cite{Torres05} for an
initially randomly aligned ensemble of room temperature D$_2$
molecules. For a particular $\Delta$$\it{t}$ and $\theta$, the
false colour-scale indicates the wavepacket probability with a
thermal average. In the range 230 $\leq$ $\Delta$$\it{t}$ $\leq$
290 fs, the ensemble is predicted to be maximally aligned with the
pump polarization direction. (b) The integrated rapid SI yield
(normalized to the theory) over the same range of $\Delta$$\it{t}$
indicates the theory describes our results with impressive
accuracy: all the major features are reproduced.}\label{fig3}
\end{figure}

In Fig. 2(b), between 200 $\leq$ $\Delta$$\it{t}$ $\leq$ 800 fs,
there is evidence of fine structure in the experimentally observed
wavepacket dependence, the result of the 1/4, 1/2 and 3/4 revivals
at respective fractions of (2$\it{Bc}$)$^{-1}$. Since the D$^+$
yield is recorded every 1 fs, an 11-point smoothing algorithm
applied to the integrated SI yield (Fig. 2b) suppresses
statistical fluctuations and hence enhances the fidelity of the
fine structure observation without loosing temporal resolution, as
shown in Fig. 2(c).

Through the Fourier transform (FT) of the experimental signal
(Fig. 2b), we recover the frequency components present in the
rotational wavepacket. Fig. 2(d) and Fig. 2(e) show histograms of
the FT of the calculation and the experimental data respectively,
the peak positions corresponding to the dominant beat frequencies
between rotational states. For a diatomic molecule, $\Delta\it{J}$
= 0, $\pm$2, thus the beat frequencies assigned to level $\it{J}$
in Fig 2 are governed by the spacing between rotational levels
$\it{J}$ and $\it{J}$+2. The peak heights are proportional to the
product of the respective quantum amplitudes. The nuclear spin
statistics of D$_2$ are also evidenced by the domination of beats
between even $\it{J}$ states over odd $\it{J}$ states.

The beats for $\it{J}$ = 0 - 2, 1 - 3 and 2 - 4 occur at 5.5 THz,
9.1 THz and 12.8 THz respectively. Rather intriguingly, the recent
work of Ergler $\it{et~al}$ \cite{Erg06}, investigating
vibrational wavepackets in the D$_2^+$ molecular ion, similar
rotational components have been observed, and attributed to the
creation of a rotational wavepacket in the molecular ion D$_2^+$.
However, the polarizability anisotropy and bond length change
dramatically under ionization from D$_2$ to D$_2^+$, with
corresponing beat frequencies being a factor of two lower than in
D$_2$. The time-domain and equivalent FT frequency domain data
described here are in excellent agreement with the expectation of
impulsively excited rotational wavepackets formed in the neutral
D$_2$ molecule.

The data presented in Fig. 2 describes only part of the rotational
dynamics of D$_2$ as those molecules outside the angular
acceptance of our detector simply are not detected. To fully
understand the emergence of rotational order, we turn now to
predicting and measuring the rapid variation of the angular
distribution of molecules around the first half-revival. Fig. 3(a)
is a false-colour representation of the expected probability
distribution as a function of evolution time $\Delta$$\it{t}$ and
$\theta$. When 250 $\leq$ $\Delta$$\it{t}$ $\leq$ 280 fs, maximal
alignment is predicted: the ensemble is expected to be
preferentially aligned in a narrow double-lobed distribution
directed along $\theta$ = 0, $\pi$ in this `quantum carpet'. By
rotating $\theta$ over the range -$\pi$ to $\pi$, the axis of
revival of the rotational wavepacket is rotated with respect to
our detector axis. The experimental data is shown in Fig. 3(b):
the relative D$^+$ yield as a function of $\theta$ and $\Delta$t
directly reflects the thermal average of the rotational wavepacket
probability over the thermal average distribution of $\it{J}$.
While some statistical scatter is present, the features of the
quantum carpet are well resolved. This demonstrates how the
molecular axes align through revival to create order. Up to
$\simeq$ 190 fs, the ensemble is to a large extent isotropic.
However in the vicinity of the half revival, the rephasing of the
wavepacket traps or confines the molecules in a well defined
spatial distribution for a short period of time. It should be
noted that the large energy spacing of the J-states in D$_2$ do
not necessitate supersonic cooling, thus this demonstration of an
ordered molecular ensemble illustrates a readily accessible
quantum system.

The mapping of this ultrafast rotational wavepacket has a number
of implications. On the most fundamental level, the observation of
the quantum carpet in D$_2$ (confirmed by our theoretical
predictions) illustrates that a room temperature ensemble of this
theoretically important molecule can be organized into an ordered
state. This can allow time- and energy-resolved measurements of
the electronic and nuclear motion with spectroscopic accuracy.
Considerable efforts are underway to perfect ultrashort ($\leq$ 5
femtoseconds) pulse generation through molecular phase modulation
\cite{Kal02}. By impulsively aligning a high pressure Raman-active
gas (short rotational period makes D$_2$ a natural choice) in a
hollow waveguide with an ultrafast pump pulse, a locally-aligned
rotational revival propagates along the waveguide at the speed of
light in the medium, and the molecular motion causes a strong
phase modulation due to the refractive index change. A second
pulse (directly equivalent to our probe) injected into the
waveguide to coincide with the rotational revival is then
spectrally broadened and compressed through the phase modulation.
Importantly, it has been demonstrated that the time dependent
phase introduced can be accurately controlled by changing the
delay between the pump and probe pulses.

The experiment was carried out at the ASTRA Laser Facility, CCLRC
Rutherford Appleton Laboratory, UK, where the assistance of J. M.
Smith, E. J. Divall, K. Ertel, O. Chekhlov, C. J. Hooker and S.
Hawkes is gratefully acknowledged. This work was funded by the
Engineering and Physical Sciences Research Council (UK). EMLE and
JW acknowledge EPSRC studentships, JMcK and CRC wish to
acknowledge funding from the Department of Employment and Learning
(NI). RT acknowledges the Spanish Department of State of Education
and Universities, and the European Social Fund.

\end{document}